\documentstyle[preprint,eqsecnum,aps]{revtex}

\begin{document}
\tighten
\draft

\title{$S$-matrix poles near the $\Lambda N$ and $\Sigma N$ 
Thresholds in the Coupled
\\$\Lambda N-\Sigma N$ System
}

\author{K. Miyagawa and H. Yamamura}

\address{
Department of Applied Physics, Okayama University of Science\\
1-1 Ridai-cho, Okayama 700, Japan
}
	    
\date{\today}
\maketitle

\begin{abstract}
We  search $t$-matrix poles for $\Lambda N-\Sigma N$ coupling
interactions using two soft core models of the Nijmegen
group which
bind the hypertriton at the correct binding energy, and 
hard core models which are still influential in hypernuclear
physics.
To treat the hard core potentials,
a useful method for calculating the off-shell $t$-matrix
is proposed. We find poles close to
the $\Sigma N$ threshold in the second or third quadrant of
the  complex plane of the  $\Sigma N$ relative momentum.
The relation between the poles and the shape of the
$\Lambda N$ elastic total cross section is discussed based
on a so-called uniformization by which two-channel $t$-matrices
become single-valued on a complex valuable.
We also find poles near the $\Lambda N$ threshold. These are
correlated to the $S$-wave $\Lambda N$ scattering lengths, the
values of which have yet to be determined.

\end{abstract}


\narrowtext

\abovedisplayskip 7mm
\belowdisplayskip 7mm
\abovedisplayshortskip 7mm
\belowdisplayshortskip 7mm
\jot 5mm   
\newfont{\myfont}{cmti12 scaled \magstep1}

\section{Introduction}

Most of our  knowledge of the $YN$ interaction  has been obtained
from heavy hypernuclei and it remains rather qualitative. 
 In contrast,
recent theoretical analyses of few-baryon systems with strangeness
such as the  three-body calculations of
$^3_\Lambda H$\cite{hyp1,hyp2,hypbnl} and the  four-body study of
$^4_\Lambda H$\cite{hiyama} 
yield more qualitative results, being
based on modern baryon-baryon forces and on rigorous solutions
of the few-body Schr\"odinger equations. Thus, they offer a great
advantage to scrutinize $YN$ interaction models. As an example, 
Refs.~\cite{hyp1,hyp2,hypbnl}
demonstrated that the Nijmegen soft core $YN$ interaction NSC89 
\cite{nsc89}
binds the hypertriton, but the J\"ulich $\tilde A$ 
potential\cite{juelich} does not.
This is possibly caused by the sizable difference between
the $^1S_0$ force
components  in the low-energy region\cite{hyp2}.
This shows that even  basic quantities such as
scattering lengths are  still undetermined.

The few-body analysis of $^3_\Lambda H$ \cite{hyp1,hyp2,hypbnl}
has also clarified
the effect of  the $\Lambda -\Sigma$
conversion which was exactly included in the coupled channel
formalism\cite{hyp1}. Although in the case of the hypertriton
the admixture of  $\Sigma NN$ states is only 0.5\%, the expectation
values of the sum of the transition potentials
$V_{\Lambda N, \Sigma N}$ and $V_{\Sigma N, \Lambda N}$
 are approximately 8\%  of  the total potential
energy, which is crucial for the binding of the hypertriton
\cite{hyp2,hypbnl}.

This knowledge about the $\Sigma$ states coupling, however, 
has been obtained from the bound state lying  below the
$\Lambda NN$  threshold.
This should  be  extended to analyses close to the $\Sigma$ threshold,
where $\Lambda -\Sigma$ conversion effects will emerge sharply. 
In this region, so-called unstable bound 
states\cite{kok}
have received attention and have been searched for experimentally.
Only one state in the $\Sigma (\Lambda) NNN$ system is confirmed in the
 reaction $^4He(K^-, \pi^-)$\cite{nagae}. The existence of such
an unstable bound state in the A=4 system was predicted by
Harada {\it et~al.} where the coupling to continuum $\Lambda NNN$ states
was approximated by a
$\Sigma N$ optical potential\cite{harada}. However, for understanding
the actual features of $\Lambda -\Sigma$ conversion, it is highly
desirable
to treat it directly using a realistic  $\Lambda N-\Sigma N$ coupling
interaction.

At present, it is technically possible to incorporate  precisely the
coupling to $\Lambda$ continuum states for only
the $\Sigma N$ and $\Sigma NN$ systems.
Afnan and Gibson\cite{afnan} calculated  the $\Lambda d$ elastic
scattering fully
incorporating this coupling but using simple phenomenological $YN$
interactions. They then found and analyzed enhancements
just below the $\Sigma NN$ threshold.
To examine more closely the $\Lambda N -\Sigma N$ coupling interaction,
a similar study applying more sophisticated meson-theoretical
interactions
is necessary. It is also important to analyze electromagnetic
hyperon-production processes \cite{tjlab}, which are experimentally
accessible.

For the $YN$ interaction, there exists a variety of strengths for
the $\Lambda N-\Sigma N$ coupling among extensively used
 meson-theoretical
 potentials. The soft core model\cite{nsc89} and the hard core model
  D\cite{nd} of the Nijmegen group show cusps in the $\Lambda N$ elastic
   total cross section just at the $\Sigma N$ threshold with different
    magnitudes, while  the Nijmegen hard core model F\cite{nf}
   and the 
   J\"ulich models\cite{juelich} show  round resonance peaks below the
   threshold. We stress here that these prominent cusps do not mean
   simple
   threshold effects, but suggest the existence of $t$-matrix poles in 
    unphysical Riemann sheets.
This is important because the poles might move and become
 unstable bound state poles if the coupling strengths  varied.
  Some examples for
 separable potentials are given in Ref.~\cite{pearce}. In this paper,
 we shall
 locate these poles and follow their trajectories 
 for the Nijmegen potentials.

Knowledge of $YN$ $t$-matrix around  the $\Sigma N$ threshold
is crucial for
the analysis of
$\Lambda NN-\Sigma NN$ continuum states  using meson-theoretical 
interactions.
This paper accordingly investigates  poles of the
$t$-matrices  for the Nijmegen F, D and the two soft core
interactions\cite{nsc89,nsc97}.
This is achieved  in momentum space and  hence  the results are directly
applicable to the  three-body calculation.  The behavior of $t$-matrices
around an inelastic threshold in coupled channel problems and the effects
of nearby poles  have  often been studied~ \cite{kok,pearce}.
In such analyses it is important to understand
the connection between various Riemann energy sheets and  how far from
the
physical region the poles are located.
In this analysis we adopt  a so-called uniformization given by
Newton\cite{newton}, by which  the $t$-matrix  for two-channel problems
becomes  single-valued  after a suitable variable is introduced in place
of energy. We thereby clearly  describe the positions and the
trajectories
of the $t$-matrix poles in the Riemann sheets.

Section II gives the expression for $\Lambda N-\Sigma N$ $t$-matrix which
is analytically continued to the complex energy  plane.  Section III
describes a method to treat a hard core potential in momentum space.
This is for the purpose of treating the Nijmegen hard core potentials
which are influential in hypernuclear physics. 
In Sec.~IV, the uniformization mentioned
above is introduced, thereby we discuss  how the shape of the $\Lambda N$
elastic total cross section around the $\Sigma N$ threshold is related to
the positions of nearby  poles.  In Sec.~V,  the positions of 
the $t$-matrix
poles  for the Nijmegen soft and hard core models are described.
We also show the trajectories of the poles when the strengths of the
potentials are increased.

\section{Analytic continuation of the $t$-matrix}

In this section, we give the expression of  the off-shell $t$-matrix for
the
$\Lambda N-\Sigma N$ system and  continue it analytically into the
complex energy plane.
 
The coupled $t$-matrices for the $\Lambda N-\Sigma N$ system are
defined by the integral equations

\begin{equation}
T_{ij}(z)=V_{ij}+\sum_k V_{ik}\, G_0^{(k)} (z)\, T_{kj}(z),
\hspace{10mm} i,j,k=1,2
\label{eq:2.1}
\end{equation}
with
\begin{equation}
G_0^{(k)} (z) = \left( z-H_0^{(k)} \right)^{-1},
\hspace{10mm} z=E+i\varepsilon
\label{eq:2.2}
\end{equation}
where $i$, $j$, $k$ have integer values of 1 and 2 for the $\Lambda N$
and $\Sigma N$
channels, respectively. The free Hamiltonian  $H_0^{(k)}$  for channel
$k$ is defined as
\begin{equation}
H_0^{(k)}=\frac{p^2_k}{2\mu_k}+m_N+m_Y^{(k)}  \, .
\label{eq:2.3}
\end{equation}
This refers to the total momentum zero frame and we denote the relative
momentum between the nucleon and the hyperon  by ${\bf  p_k}$ and 
the reduced mass of  channel $k$  by $\mu_k$. The masses
 $m_Y^{(k)}$$(k=1,2)$ indicate $m_\Lambda$ and $m_\Sigma$ respectively.
After performing the partial-wave decomposition in momentum space, we
express the
projected $t$-matrix elements  for a given total angular momentum and
parity again by $T$. Then Eq.~(\ref{eq:2.1})  reads
\begin{eqnarray}
< p \,|\, T_{ij}(z) \,|\, p^\prime >&=&< p \,|\, V_{ij} \,|\, p^\prime >
\nonumber \\
&+&\sum_k \int_0^\infty dp^{\prime\prime} {p^{\prime\prime}}^2 < p \,|\,
V_{ik} \,|\, p^{\prime\prime} >
\frac{1}{e_k - \frac{{p^{\prime\prime}}^2}{2\mu_k} + i\varepsilon }
< p^{\prime\prime} |\, T_{kj}(z) |\, p^{\prime} >
\label{eq:2.4}
\end{eqnarray}
with
\begin{equation}
e_k\equiv  \frac{q_k^2}{2\mu_k}=
E-m_N-m_Y^{(k)} \,\, .
\label{eq:2.5}
\end{equation}
To simplify the notation, the $p$ indices have been omitted, and
the  partial-wave  elements are assumed to
have  no coupling between different orbital angular momenta or 
channel-spin states. The  extension to the case with
couplings is straightforward.

Now consider the  energy $E$ to be a complex number. Hence
$e_k$ and
\begin{equation}
q_k =\sqrt{2\mu_k e_k}
\label{eq:2.6}
\end{equation}
are  complex numbers. We introduce the function
\begin{equation}
h_k(p^{\prime\prime})\equiv  < p \,|\, V_{ik} \,|\, p^{\prime \prime}
> < p^{\prime\prime} \,|\, T_{kj}(z) \,|\, p^{\prime} >
\label{eq:2.7}
\end{equation}
and define each term in the $k-$summation of the right-hand side of
Eq.~(\ref{eq:2.4}) by $I_k(e_k)$ as
\begin{equation}
I_k(e_k)= \int^\infty_0 dp^{\prime\prime}\,\,
                     \frac{{p^{\prime\prime}}^2 h_k(p^{\prime\prime})}
                  {e_k -\frac{{p^{\prime\prime}}^2}{2\mu_k}} \,\,\,  .
\label{eq:2.8}
\end{equation}
This function has a cut for $e_k \geq 0$, in other words,
a cut along $m_N+m_Y^{(k)}\leq E <\infty$.
Thus, there are two cuts in the $E$ plane starting at the  $N+\Lambda$
 and $N+\Sigma$ thresholds, respectively.
The function values beyond the cuts are defined by analytic continuation.
 This is achieved by modifying Eq.~(\ref{eq:2.8}) as
\begin{equation}
I_k(e_k)=
 \int^{\infty}_0 dp^{\prime\prime}\
\frac{ {p^{\prime\prime}}^2 h_k( p^{\prime\prime} ) 
- 2\mu_k e_k h_k(q_k) }
{ e_k - \frac{{p^{\prime\prime}}^2}{2\mu_k} }
+ h_k (q_k) \int^{\infty}_0 dp^{\prime\prime}
          \frac{2\mu_k e_k }{e_k -\frac{{p^{\prime\prime}}^2}{2\mu_k} }
\label{eq:2.9}
\end{equation}
where we assume that $h_k(p)$ can be continued analytically
and has no singularity in the trajectory from real $p$ to the
complex value
$q_k$ given in Eq.~(\ref{eq:2.6}).
This is true for the case here. The cut now appears explicitly
in
the second term of Eq.~(\ref{eq:2.9}).
It is then easy to show
\begin{equation}
\int^{\infty}_0 dp^{\prime\prime}
     \frac{2\mu_k e_k}{e_k -\frac{{p^{\prime\prime}}^2}{2\mu_k} }\ =
  -i\pi \mu_k \sqrt{2\mu_k e_k}
  = -i\pi \mu_k q_k
\label{eq:2.10}
\end{equation}
which defines the integral in both sheets of the Riemann $e_k$ surface,
corresponding to positive
and negative imaginary parts of $q_k$.
From Eqs.~(\ref{eq:2.9}) and (\ref{eq:2.10}), we can
rewrite  Eq.~(\ref{eq:2.4}) as
\begin{eqnarray}
\lefteqn{< p \,|\, T_{ij} (z) \,|\, p^{\prime} >
=< p \,|\, V_{ij} \,|\, p^\prime > } \nonumber \\
&+&
\sum_k \left[ \, \int^\infty_0 dp^{\prime\prime}
\left(
\frac{{p^{\prime\prime}}^2
 < p \,|\, V_{ik} \,|\, p^{\prime\prime} >
<p^{\prime\prime}\,|\, T_{kj}(z) \,|\, p^\prime >}
{e_k-\frac{{p^{\prime\prime}}^2}{2\mu_k}}
\right. \right. \nonumber \\
&-&\left. \left.
 \frac{ 2\mu_k e_k < p \,|\, V_{ik} \,|\, q_k >
< q_k \,|\, T_{kj} (z) \,|\, p^\prime >}  
{e_k-\frac{{p^{\prime\prime}}^2}{2\mu_k}}
 \right)
-i\pi \, \mu_k q_k < p \,|\, V_{ik} \,|\, q_k >
< q_k \,|\, T_{kj}(z) \,|\, p^{\prime} > \right]  .
 \nonumber \\ &&
\label{eq:2.11}
\end{eqnarray}
This equation contains
a new $t$-matrix element
$<q_k \mid T_{kj} \mid p'>$ which requires the additional
equation
\samepage{
\begin{eqnarray}
\lefteqn{< q_k \,|\, T_{ij} (z) \,|\, p^{\prime} >
=< q_k \,|\, V_{ij} \,|\, p^\prime >} \nonumber \\
&+&\sum_k \left[ \, \int^\infty_0 dp^{\prime\prime}
\left( \frac{{p^{\prime\prime}}^2
        < q_k \,|\, V_{ik} \,|\, p^{\prime\prime} >
        <p^{\prime\prime}\,|\, T_{kj}(z) \,|\, p^\prime >
  }
{e_k-\frac{{p^{\prime\prime}}^2}{2\mu_k}}
\right. \right. \nonumber \\
&-&\left. \left.
\frac{2\mu_k e_k
      < q_k \,|\, V_{ik} \,|\, q_k >
      < q_k \,|\, T_{kj} (z) \,|\, p^\prime >
       }
    {e_k-\frac{{p^{\prime\prime}}^2}{2\mu_k}}
\right)
-i\pi \, \mu_k q_k < q_k \,|\, V_{ik} \,|\, q_k >
< q_k \,|\, T_{kj}(z) \,|\, p^\prime > \right]  .
\nonumber \\
&&
\label{eq:2.12}
\end{eqnarray}
}
The Eqs.~(\ref{eq:2.11}) and (\ref{eq:2.12})  form a closed set of
integral equations\cite{gtext} and define the $t$-matrix elements
on the entire $q_k$ planes or the Riemann surface  of the complex
 energy $E$.
This set is solved in the following sections.

\section{$t$-matrix for a hard core potential}

Although it is now rare to represent the short-range
repulsion of the NN interaction by a hard core, the Nijmegen D and F
models
of  the $YN$ interaction with hard cores are still used frequently 
in hypernuclear physics. Therefore, in this section
we explore a method to obtain  the off-shell $t$-matrix
for a hard core potential in momentum space.

The off-shell $t$-matrix can be expressed  as
\begin{equation}
< \vec{p} \,|\, T(z) \,|\, \vec{k} >
=< \vec{p} \,|\, V \,|\, \Psi_{q,\, \vec{k}}^{(+)} >
\label{eq:3.1}
\end{equation}
where $\Psi_{q, \, \vec k}^{(+)} $ is defined by
\begin{equation}
|\, \Psi_{ q,\, \vec{k}}^{(+)}  > = |\, \vec{k} >
+ \, G_0(z) \, V \,|\, \Psi_{ q,\, \vec{k}} ^{(+)} >
\label{eq:3.2}
\end{equation}
with
\begin{equation}
z=\frac{q^2}{2\mu}+i\varepsilon \,\,\, .
\label{eq:3.3}
\end{equation}
To simplify the notation, the $\Lambda N-\Sigma N$ coupling
has been omitted.
First, we divide the interaction $V$ into the pure hard core part
 $U$ and the remainder $\hat V$ as
\begin{equation}
V=U+\hat{V}
\label{eq:3.4}
\end{equation}
and use the two-potential formula$\cite{gw}$ to obtain 
\begin{equation}
< \vec{p} \,|\, T(z) \,|\, \vec{k} >
= < \vec{p} \,|\, U \,|\, \Phi^{(+)}_{ q,\, \vec{k}} >
+ < \Phi^{(-)}_{ q,\, \vec{p}} \,|\, \hat{V} \,|\,
\Psi_{ q,\, \vec{k}}^{(+)} >
\label{eq:3.5}
\end{equation}
with
\begin{equation}
|\, \Phi_{ q,\, \vec{k}}^{(+)} >\, = |\, \vec{k} > +
\,G_0(z)\, U \,|\, \Phi_{ q,\, \vec{k}}^{(+)} > \,\,  ,
\label{eq:3.6}
\end{equation}
\begin{equation}
|\, \Phi_{ q,\, \vec{p}}^{(-)} >\, = |\, \vec{p} >
+ \, G_0(z^{\ast}) \,U \,|\, \Phi_{ q,\, \vec{p}}^{(-)} > \,\,  .
\label{eq:3.7}
\end{equation}
As we shall show later,  the first term of the right-hand side of 
Eq.~(\ref{eq:3.5}) is expressed analytically, and  the second term
satisfies an integral equation similar to the Lippmann-Schwinger equation
 which can be solved using a standard method.
Our method is thus a natural extension of a standard treatment without a
hard
core, and  is therefore useful not only for the present purpose, but also
for
 other few-body calculations in momentum space.

The analytic expression of the first term in Eq.~(\ref{eq:3.5})
has already been given 
by Takemiya$\cite{takemiya}$, who  proposed a method to evaluate
the  off-shell $t$-matrix for a hard-core potential in coordinate space.
Here, we use this method only in the treatment of the pure hard-core part
of the  formula. 

$\Phi_{ q,\, \vec{k}}^{(+)}$ and
$\Phi_{ q,\, \vec{p}}^{(-)}$ in 
Eqs.~(\ref{eq:3.6}) and  (\ref{eq:3.7})  can be expanded into  partial
waves
\begin{equation}
\Phi^{(\pm)}_{ q,\, \vec{k}}(\vec{r})
=\sum_{
\stackrel{\scriptstyle l^\prime s^\prime}{\stackrel{\scriptstyle ls}{JM}}
}
 {\mbox{\myfont y}}^{JM}_{l^\prime s^\prime} (\hat{r})
\,{\Phi^{J(\pm)}_{l^\prime s^\prime ls}}(q,k,r) \,
{{\mbox{\myfont y}}_{ls}^{JM}}^{\dag} (\hat{k})
\label{eq:3.8}
\end{equation}
and  similarly $\Psi_{ q,\, \vec{k}}^{(+)}$.
Here, $ {\mbox{\myfont y}}^{JM}_{L S}$ is the
simultaneous eigenfunction of $L^2$, $S^2$, $ J^2$ and  $ J_z$. 
We denote the pure-hard core part of  Eq.~(\ref{eq:3.5}) as
\begin{equation}
< \vec{p} \,|\, \tilde{t}(z) \,|\, \vec{k} >
\equiv < \vec{p} \,| \,U\, |\, \Phi^{(+)}_{ q,\, \vec{k}} >
\label{eq:3.9}
\end{equation}
and decompose it into partial waves
\begin{equation}
< \vec{p} \,|\, \tilde{t} (z) \,|\, \vec{k} >
=\sum_{
\stackrel{\scriptstyle l^\prime s^\prime}{\stackrel{\scriptstyle ls}{JM}}
} {\mbox{\myfont y}}^{JM}_{l^\prime s^\prime} (\hat{r}) \,
\tilde{t}_{l^\prime s^\prime ls}^J (p,k;z) \,
{ {\mbox{\myfont y}}_{ls}^{JM}}^{\dag}(\hat{k}) \,\, .
\label{eq:3.10}
\end{equation}
Reference~\cite{takemiya} proves that if we introduce a function $\chi$
defined by
\begin{equation}
\frac{ {\chi_{l^\prime s^\prime ls}^{J(\pm)}}(q,k,r) }{r}
\sqrt{ \frac{2}{\pi} } \:i^l
\equiv {\Phi^{J(\pm)}_{l^\prime s^\prime ls}}(q,k,r)-
\sqrt{ \frac{2}{\pi} } \:i^l j_l(kr)
\label{eq:3.11}
\end{equation}
it satisfies the equation
\begin{equation}
\left( q^2 + \frac{d^2}{d r^2} - \frac{l^\prime (l^\prime + 1)}{r^2}
\right)
{\chi^{J(\pm)}_{l^\prime s^\prime ls}}(q,k,r)
-\sum_{ l^{\prime\prime} s^{\prime\prime}
       }
{2\mu} \, U^{J}_{l^\prime s^\prime l^{\prime\prime}
s^{\prime\prime}} (r)
\, {\chi^{J(\pm)}_{l^{\prime\prime} s^{\prime\prime} ls}}(q,k,r)
=r{2\mu} \, U^J_{l^{\prime} s^{\prime} ls}(r) \, j_l(kr)
\label{eq:3.12}
\end{equation}
and  the off-shell element of $\tilde t$ is given by $\chi$ as
\begin{equation}
\tilde{t}^J_{l^{\prime} s^{\prime} ls}(p,k;z)=\frac{1}{2\mu}\,
\frac{2}{\pi} \, i^{-l^{\prime}+l}
\int^{\infty}_0 dr \, r \, j_{l^{\prime}} (pr)
\left( q^2+\frac{d^2}{dr^2}
-\frac{l^{\prime}(l^{\prime}+1)}{r^2}\right) {\chi^{J(+)}_{l^{\prime}
s^{\prime} ls}} (q,k,r) \,\, .
\label{eq:3.13}
\end{equation}
Following the method described in Ref.~\cite{takemiya}
one arrives at the analytic
 expression of the $\tilde t$ element in Eq.~(\ref{eq:3.13}).
 For a pure hard core potential with radius $c$,
 Eq.~(\ref{eq:3.12}) has the solution 
\begin{equation}
{\chi^{J(\pm)}_{l^{\prime} s^{\prime} ls}}(q,k,r) = \delta_{l^{\prime} l}
\delta_{s^{\prime} s} \times
\left\{
\begin{array}{ll}
\begin{minipage}{0.3\textwidth}
\[-r\,j_l (kr)\]
\end{minipage}
&(r\le c) \\
\begin{minipage}{0.3\textwidth}
\[
-\frac{j_l(kc)}{h_l^{(\pm)}(qc)} \,r\,h_l^{(\pm)} (qr)
\]
\end{minipage}
&(r\ge c)
\end{array}
\right.
\label{eq:3.14}
\end{equation}
Note, there is no coupling in $\ell$ and $s$, a result
which can be found by generating the hard-core potential matrix $U$
as limits of square well potentials (see Ref.~\cite{takemiya}).
Further, from   Eq.~(\ref{eq:3.14}) the integration in
 Eq.~(\ref{eq:3.13}) is limited to $r\leq c$
and performing the integration  we obtain the final expression  for
$\tilde t$
\begin{eqnarray}
\tilde{t}^J_{l^{\prime} s^{\prime} ls}(p,k;z)&=&\delta_{l^{\prime} l}
\delta_{s^{\prime} s}
 \, \frac{1}{2\mu} \, \frac{2}{\pi} \, i^{-l^{\prime} + l}
 \nonumber \\
&\times&\left[ \left. -\,c \, j_l (pc) \frac{d}{dr} \,r\, h_l^{(+)} (qr)
\right|_{r=c}
 \, \frac{j_l(kc)}{h_l^{(+)}(qc)}+ \left. \frac{d}{dr} \, r\, j_l(pr)
 \right|_{r=c} \,c \:j_l (kc)
 \right.
\nonumber \\
&-& \left. \left( q^2-p^2 \right) \int^c_0 dr\, r^2 \, j_l (pr)
\,j_l (kr)
\right]  .
\label{eq:3.15}
\end{eqnarray}
The last term on the right-hand side of this equation is shown
 in Ref.~\cite{takemiya} to be
\begin{equation}
\int^c_0 dr \,r^2 \,j_l(pr) \,j_l(kr)
=\left\{
\begin{array}{ll}  
\begin{minipage}{0.6\textwidth}
\[\frac{c^2}{k^2-p^2} \left( \,k\, j_l(pc) \,j_{l+1}(kc)-p \,
j_l (kc) \,j_{l+1} (pc) \right)\]
\end{minipage}
& (p\neq k) \\
\begin{minipage}{0.6\textwidth}
\[
\frac{c^2}{2k} \left( \,k\,c \left( \,j_l^2 (kc) + j_{l+1}^2 (kc)
\,\right)
 - (2l+1)\,j_l(kc)\,
j_{l+1}(kc)
\right)\]
\end{minipage}
& (p=k)
\end{array}
\right.
\label{eq:3.16}
\end{equation}

Next, let us consider the second part of the two-potential
formula~(\ref{eq:3.5}). The state $\Psi_{ q,\, \vec k}^{(+)} $ given
 in Eq.~(\ref{eq:3.2}) satisfies another equation\cite{gw}
\begin{equation}
| \,\Psi_{ q,\, \vec{k}}^{(+)}  > \,= |\, \Phi_{q,\, \vec{k}}^{(+)} >
 + \,G_U (z) \,\hat{V} \,|\, \Psi_{ q,\, \vec{k}}^{(+)}  >
\label{eq:3.17}
\end{equation}
with
\begin{equation}
G_U(z) = G_0 (z) + G_0 (z) \,U\, G_U (z) \,\, .
\label{eq:3.18}
\end{equation}
Hence, if we define $\hat t$ by
\begin{equation}
< \vec{p} \,|\, \hat{t} (z) \,|\, \vec{k} >\,
\equiv \,< \Phi_{ q,\, \vec{p}}^{(-)} \,|\, \hat{V} \,|\,
 \Psi_{ q,\, \vec{k}}^{(+)}  >
\label{eq:3.19}
\end{equation}
then the second part of the two-potential formula becomes
\begin{equation}
< \vec{p} \,|\, \hat{t}(z) \,|\, \vec{k} >
\,=\, < \Phi_{ q,\, \vec{p}}^{(-)} \,|\, \hat{V} \,|\,
 \Phi_{ q,\, \vec{k}}^{(+)} >
+ < \Phi_{ q,\, \vec{p}}^{(-)} \,|\, \hat{V} \,G_U (z) \,\hat{V} \,|\,
 \Psi_{ q,\, \vec{k}}^{(+)}  >  \, .
\label{eq:3.20}
\end{equation}
Observing that
\begin{eqnarray}
< \vec{p}^{\:\prime} \,|\, G_U &=& <\vec{p}^{\:\prime} \,|\, G_0
 \left(\, 1+UG_U \,\right) \nonumber \\
&=& \frac{1}{z-\frac{{p^{\prime}}^2}{2\mu}}
<\Phi_{q, \, \vec{p}^{\:\prime}}^{(-)} |
\label{eq:3.21}
\end{eqnarray}
and applying it to the second term of the right-hand side of
Eq.~(\ref{eq:3.20}), we arrive at the integral equation for $\hat t$
\begin{equation}
<\vec{p}\,|\, \hat{t}(z) \,|\,\vec{k}>=< \Phi_{ q,\, \vec{p}}^{(-)}\,|\,
 \hat{V} \,|
\, \Phi_{ q,\, \vec{k}}^{(+)} >
+\int d\vec{p}^{\:\prime} < \Phi_{ q,\, \vec{p}}^{(-)} \,|\, \hat{V} 
\,|\,
 \vec{p}^{\:\prime} >
\frac{1}{z-\frac{{p^{\prime}}^2}{2\mu}} < \vec{p}^{\:\prime}
 \,|\, \hat{t} (z) \,|\, \vec{k} > \, .
\label{eq:3.22}
\end{equation}

The inputs to this integral equation,
$<\Phi^{(-)}_{ q,\, \vec p} \mid \hat V \mid \Phi^{(+)}_{ q,\, \vec k}>$
and
$<\Phi^{(-)}_{ q,\, \vec p} \mid \hat V \mid \vec p\, ' >$,
are expressed by the scattering states from the pure hard-core part,
  $\Phi^{(\pm)}$, and
the remainder of the potential, $\hat V$.
From Eqs.~(\ref{eq:3.11}) and (\ref{eq:3.14}), the scattering
states
$\Phi^{(\pm )}$ can be
 expressed simply by spherical Bessel and Hankel functions as
\begin{equation}
{\Phi^{J(\pm)}_{l^{\prime} s^{\prime} ls}} (q,k,r)= \left\{
\begin{array}{ll}
\begin{minipage}{0.5\textwidth}
\[0\]
\end{minipage}
&(r<c)\\
\begin{minipage}{0.5\textwidth}
\[
\delta_{l^{\prime} l} \delta_{s^{\prime} s} \,\sqrt{\frac{2}{\pi}} \:i^l
\left( \,j_l(kr) - \frac{j_l(kc)}{h_l^{(\pm)}(qc)}
\,h_l^{(\pm)}(qr) \,\right)
\]
\end{minipage}
&(r>c)
\end{array}
\right.
\label{eq:3.23}
\end{equation}
allowing the inputs to be easily calculated.
Thus, the integral equation~(\ref{eq:3.23})  can be solved in a similar
manner
as described in Sec.~II.
							 
 Consequently, combining the analytic expression given in
 Eqs.~(\ref{eq:3.15}) and (\ref{eq:3.16}) with the solution of this 
integral  equation, we easily obtain the off-shell $t$-matrix
for a hard-core potential.

\section{Cusps and round peaks caused by nearby poles}

As described in Sec.~I, the main aim of this paper is to search
 $t$-matrix  poles for various $YN$ interactions around the $\Sigma N$
 threshold. In Eq.~(\ref{eq:2.4}), the $t$-matrix elements are defined
  by the relative momenta between the hyperons and the nucleon,
   $q_1$ in the case of $\Lambda$-N and $q_2$ in the case of $\Sigma$-N.
However, these momenta are
not independent and  are related to the energy $E$ through
Eq.~(\ref{eq:2.5}).
This can be rewritten as
\begin{equation}
\frac{q^2_1}{2\mu_1}+m_N+m_{\Lambda}=\frac{q^2_2}{2\mu_2}
+m_N+m_{\Sigma}=E \,\, .
\label{eq:4.1}
\end{equation}
Thus, each $t$-matrix element is a function of the energy $E$, and has
 branch points at  the two thresholds
$E=m_N+m_\Lambda$ and $E=m_N+m_\Sigma$.
We therefore encounter a somewhat complicated Riemann energy surface
with four sheets, and must specify how they are related
to the upper and lower halves of the $q_1$ and $q_2$ planes
\cite{kok,pearce}.
In two-channel problems,  a procedure called
 uniformization\cite{newton}  is very convenient to map
 the 4 Riemann sheets into one plane. This is used
 in the present analysis.  The uniformization  procedure
  introduces a new variable in place of the energy, in terms of which
  the $t$-matrix becomes single-valued.
 Following Ref.~\cite{newton}, we introduce such  a
 variable $\omega$ which satisfies
\begin{equation}
\frac{q_1}{\sqrt{2\mu_1}} + \frac{q_2}{\sqrt{2\mu_2}} =
\Delta\  \omega
\label{eq:4.2x}
\end{equation}
and
\begin{equation}
\frac{q_1}{\sqrt{2\mu_1}} - \frac{q_2}{\sqrt{2\mu_2}} = \Delta\ 
\omega^{-1}
\label{eq:4.2}
\end{equation}
with
\begin{equation}
\Delta^2 \equiv m_{\Sigma}-m_{\Lambda} \,\, .
\label{eq:4.3}
\end{equation}
By these relations~(\ref{eq:4.2}) and (\ref{eq:4.2x}) it is easy to
realize Eq.~(\ref{eq:4.1}).
These equations constitute a mapping of the Riemann energy surface to the
complex $\omega$ plane which is shown in Fig.~1.
Of course, there are 4 possible quadrants  where  $q_1$ can be located,
and for each $q_1$  two different values of $q_2$ are allowed by
Eq.~(\ref{eq:4.1}).
Hence, there are 8 possible cases in specifying to which quadrants both
$q_1$ and $q_2$ belong  on their own complex planes. The complex $\omega$
plane in Fig.~1 is divided accordingly into 8 parts, each of which
contains two numbers inside square brackets  indicating the quadrants
to which $q_1$ and $q_2$ belong.
The bold line expresses  the region where  bound or scattering states
exist if present.
The $\Lambda N$ threshold is located  at $\omega=i$, and the
$\Sigma N$ threshold resides  at $\omega=1$. If moving
counter-clockwise
around $\omega=i$,  the quadrant  to which $q_1$ belongs  changes as
$1\rightarrow 2\rightarrow 3\rightarrow 4$, and at the same time  the
 quadrant of $q_2$ changes as
 $1\rightarrow 2\rightarrow 1\rightarrow 2$. 
On the other hand, if one moves  around $\omega=1$ which corresponds to
the
$\Sigma N$ threshold, the quadrant  to which $q_2$ belongs  varies  as
$1\rightarrow 2\rightarrow 3\rightarrow 4$
and the quadrants of $q_1$  as 
$1\rightarrow 4\rightarrow 1\rightarrow 4$.

Let us now consider  the relation between the shapes of the $\Lambda N$
elastic  total cross section and the  positions of a pole near the
$\Sigma N$ threshold.
 One important difference to single channel problems  is that there
exists the
 region [1,3] touching the  $\Sigma N$ threshold.
Suppose a pole exists in this region close to the threshold, then the
$\Lambda N$  elastic total cross section takes the shape of a cusp just
at the threshold.
On the other hand, if a pole resides in the region [4,2] or [4,4] close
to
the bold line mentioned above, the cross section shows a round peak of
the Breit-Wigner form.
A pole lying in the region [4,2] is often called  an unstable bound
state (UBS) pole\cite{kok}. 

We shall now discuss the above mentioned behavior
of the cross sections. Assuming that the $t$-matrix has a
pole at the position $(q_1,q_2)=(\alpha_1,\alpha_2)$ and the
corresponding energy is $E_0$, it follows that      
\begin{equation}
E_0 = \frac{\alpha_1^2}{2\mu_1}+m_N+m_{\Lambda} = \frac{\alpha_2^2}
{2\mu_2}+m_N+m_{\Sigma} \,\, .
\label{eq:4.4}
\end{equation}
Then, the $t$-matrix elements around the pole  
can be approximated as
\begin{equation}
<p\,|\,T_{ij}(E)\,|\,p^{\prime}> \,\simeq\, 
\frac{R_{ij}(p,p^{\prime})}{E-E_0}  \, .
\label{eq:4.5}
\end{equation}
The approximation (\ref{eq:4.5}) of a first-order pole   holds even
 in the case when the pole resides in  Riemann sheets other 
 than the first,
 which is  proved  for example
in Ref.~\cite{elster}  for a single channel case.
The extension to the coupled $\Lambda N-\Sigma N$ system is 
straightforward.
Notice, however, for all pairs of $(q_1,q_2)=(\pm\alpha_1,\pm\alpha_2)$
the energies defined by Eq.~(\ref{eq:4.1}) take  the same value  $E_0$
but reside in different places on the Riemann energy surface.
The approximation (\ref{eq:4.5}) is therefore valid
only around    $(q_1,q_2)=(\alpha_1,\alpha_2)$.

From Eqs.~(\ref{eq:4.1}) and (\ref{eq:4.4}), we can rewrite
 Eq.~(\ref{eq:4.5}) as
\begin{equation}
<p\,|\,T_{ij}(E)\,|\,p^{\prime}> \,\simeq\, \frac{ \tilde{R}_{ij}
(p,p^{\prime}) }{q_1-\alpha_1}
\label{eq:4.6}
\end{equation}
or
\begin{equation}
<p\,|\,T_{ij}(E)\,|\,p^{\prime}> \,\simeq\,
\frac{ \bar{R}_{ij} (p,p^{\prime}) }
{q_2-\alpha_2} \,\, .
\label{eq:4.7}
\end{equation}
However, the expression (\ref{eq:4.6}) is not appropriate  around the
$\Sigma N$ threshold. As already mentioned, this is because if the pole
moves around the $\Sigma N$ threshold, the quadrant to which
$\alpha_1$ belongs changes as
$1\rightarrow 4\rightarrow 1\rightarrow 4$,
while the quadrant in which $\alpha_2$
 is located changes as  $1\rightarrow 2\rightarrow 3\rightarrow 4$.
Therefore, the expression  (\ref{eq:4.6}) can not distinguish whether
the pole is situated  in the regions [4,2] or [4,4],
or in the regions [1,1] or [1,3].
On the other hand, the expression  (\ref{eq:4.7})
can distinguish between the regions and so will be used here.

Let us now infer from Eq.~(\ref{eq:4.7})  the shapes of the $\Lambda N$
elastic
total cross section $\sigma$ around the $\Sigma N$ threshold. Since
\begin{equation}
\sigma \propto \left| \,<q_1\,|\,T_{11}(E)\,|\,q_1> \,\right|^2
\label{eq:4.8}
\end{equation}
the  $q_2$ dependence of the cross section becomes roughly
\begin{equation}
\sigma \propto \left| \frac{1}{q_2-\alpha_2} \right|^2  \, .
\label{eq:4.9}
\end{equation}
Writing $\alpha_2 =a+ib$, we have
\begin{equation}
\sigma \propto \left\{
\begin{array}{ll}
\begin{minipage}{0.25\textwidth}
\[
\frac{1}{\left( |q_2| - b \right)^2 + a^2}
\]
\end{minipage}
&(\ q_2=i|q_2|:\ {\rm below\ the\ \Sigma N\ threshold}\ )
\\
\begin{minipage}{0.25\textwidth}
\[
\frac{1}{\left( q_2 - a \right)^2 + b^2}
\]
\end{minipage}
&(\ q_2>0:\ {\rm above\ the\ \Sigma N\ threshold})
\end{array}
\right.
\label{eq:4.10}
\end{equation}

For the three cases, when the pole is located in the regions [4,2],
 [1,3] and [4,4], we  plot  the cross sections $\sigma$ expressed by
Eq.~(\ref{eq:4.10}) in Figs.~2, 3 and 4, respectively.
Notice that Figs.~2(c), 3(c) and 4(c)  show the cross section as a
function of the energy $E$, hence its derivative  at the threshold
energy is infinite
according to the relation~(\ref{eq:4.1}). If the pole is located in the
regions [4,2] or [4,4], the cross sections show
round peaks of the Breit-Wigner form (Figs.~2(c) or 4(c)), and
are quite similar to the resonances in single channel problems.
In contrast, if the pole sits in the region [1,3], the cross section
forms
a large cusp just  at the threshold (Fig.~3(c)). In Ref.~\cite{kok},
these types of poles are
named  inelastic  virtual state  poles. We should recognize that such
a large
cusp is  caused by the pole, and is not a simple threshold effect.
Some such poles, as we shall show in the next section,  can actually
move into the region [4,2] and convert to unstable bound state
poles
when the potential strength is slightly increased.

\section{Results and discussions}

We searched  $t$-matrix poles for various meson theoretical $YN$ 
interactions
in the manner described in Secs.~II and III. 
 We used two soft core
models of the Nijmegen group, NSC89 \cite{nsc89} and
the recently  proposed  new soft core model NSC97 \cite{nsc97}, which
includes
  six different versions  named a, b, c, d, e and f. In this study we
  analyzed
  NSC97f.  Both soft core models NSC89 and NSC97f reproduce the
  correct binding energy of the hypertriton\cite{hyp2,hypbnl,miyagawa}.
We also chose hard core models D\cite{nd}  and
F\cite{nf} of the Nijmegen group (abbreviated as ND and NF respectively)
which
are still used  in hypernuclear physics studies.

Fig.~5 shows the $\Lambda N$ elastic total cross sections
around the $\Sigma N$ threshold for the force
models above.  The model NF yields a round peak, while
   ND and NSC89 form cusps just at the threshold.
    For NSC97f, the shape is unclear. All the
 enhancements are found to be caused by the $^3S_1-^3D_1$
force component.
Unfortunately,
there exist only sparse experimental data
of the $\Lambda N$ cross sections,  and so we can not determine
its actual shape.
However, a prominent peak around the $\Sigma N$ threshold has
been  observed
in the  $K^-+d\rightarrow p+\Lambda+\pi^-$ reaction\cite{kd}.

For every potential used, we found a pole near the $\Sigma N$ threshold
in the $^3S_1-^3D_1$ wave. These are shown in Table~I.
In Fig.~6, the poles are also displayed in the complex $q_2$ ($\Sigma -N$
relative momentum) plane. For NSC97f and NF, the poles are located
 in the
[4,2] region of the $\omega$ plane, and for NSC89 and ND, they lie
 in the region [1,3].
The relation between the position of the pole and the shape of the
$\Lambda-N$
cross section described in Sec.~IV holds for all potentials except
NSC97f. The pole for NSC97f is close to
the boundary between
the regions [4,2] and [1,3], and it is farther from the
imaginary axis of the $q_2$ plane than for NF. This explains
why
the shape of the $\Lambda N$ cross sections for this
potential  is not a definite example of a cusp or a round peak type.

For all the interaction models,
the poles are close to the $\Sigma N$ threshold and cause some
enhancements.  For NSC97f, the unstable bound state exists
in the two-body $YN$ system, and  very likely in the $YNN$ system.
We should emphasize that the poles in the region
[1,3] which produce the cusps are as equally important as those in the
region
[4,2]. To demonstrate this, we calculated the
trajectory of the pole for the potential ND, multiplying it  by an
overall
  strength
 parameter  $\lambda$. The trajectory is shown in Fig~7.
 The pole moves from the region [1,3] into [4,2] as the potential
strength
 increases, and becomes an unstable bound state pole.
 As for the location of poles in the complex energy sheets,
we refer the readers to Ref.~\cite{kok}
where  they are nicely illustrated.

We discovered that poles also exist near the $\Lambda N$
threshold.   In Table~II, the antibound-state poles below the 
$\Lambda N$ threshold
are shown for the $^1S_0$ and $^3S_1-^3D_1$ waves. The $^1S_0$ poles are
relatively close to the threshold, and as expected correlate to the
scattering length.
As mentioned earlier, the $\Lambda N$ scattering lengths
have yet to be determined
because of scant cross section data.
However, the analyses of the hypertriton\cite{hyp2,hypbnl,miyagawa}
 constrain the $S$-wave scattering lengths.
 The potentials NSC97f and NSC89
 which reproduce both the hypertriton binding energy and the
 $\Lambda N$
 cross section data have a $^1S_0$ scattering length within
  $-2.6$ to $-2.4$ fm, and a $^3S_1$ scattering length
 within  $-1.7$ to $-1.3$ fm.
 The corresponding position of the  $^1S_0$ pole is at about
   $-0.27 i$ fm$^{-1}$
 in the $q_1$($\Lambda -N$ relative momentum) complex plane.

 Finally, we would like to point out that
the analyses of the kaon photoproduction
 processes, $d(\gamma,\, K^+)YN$ or $^3$He$(\gamma,\, K^+)YNN$
offer a  very  promising way to clarify the effects caused by
the $YN$ final-state
interaction around the $\Lambda N$ and $\Sigma N$ thresholds.
These processes are  experimentally feasible at
TJLAB and SPring-8.
Further, the  interactions of the photon and
$K^+$ meson with the baryons
are comparatively weak, which enables one to formulate and
calculate these reactions rather well. All the techniques and
insights gained in this article are
immediately applicable to those reactions and we plan to
perform such calculations in the near future.

\acknowledgements
We would like to thank W.~Gl\"ockle for discussions, and
critical and helpful comments.
We are also grateful to L.~Anthony for proofreading
the final manuscript.


\pagebreak

 \begin{table}
 \caption{ Poles near the $\Sigma N$ threshold for the
 component ${}^3S_1-{}^3D_1$ of the various force models.
 The positions of the poles  are shown on the complex planes of the
 relative momenta in the $\Lambda N$ and $\Sigma N$ channels,
 $q_1$ and $q_2$, respectively.
 The corresponding center-of-mass energies are indicated by $E$.}
 \begin{tabular}{cccc}
model&$q_1\,(\mbox{fm}^{-1})$&$q_2 \,(\mbox{fm}^{-1})$&$E$
 (MeV)\\
 \tableline 
NSC97f&$(1.46, -0.04)$&$(-0.35, 0.15)$&$(2135.6, -3.89)$\\
NSC89&$(1.37, 0.01)$&$(-0.04, -0.39)$&$(2126.3, 1.07)$\\
ND&$(1.43, 0.01)$&$(-0.18, -0.08)$&$(2132.8, 1.07)$\\
NF&$(1.44, -0.02)$&$(-0.28, 0.12)$&$(2134.2, -2.49$)\\
 \end{tabular}
 \end{table}
 \begin{table}
 \caption{ Poles below the $\Lambda N$ threshold. The scattering
 lengths
 indicated by $a$ are also shown. See the caption to Table I for other
  details.}
 \begin{tabular}{cccccc}
model&partial wave&$q_1\,(\mbox{fm}^{-1})$&$q_2 \,
(\mbox{fm}^{-1})$&$E$ (MeV)&$a$ (fm)\\
 \tableline 
NSC97f&${}^1S_0$&$(0, -0.27)$&$(0, 1.47)$&$(2051.8, 0)$&$-2.59$\\
&${}^3S_1-{}^3D_1$&$(0, -0.37)$&$(0, 1.49)$&$(2049.3, 0)$&$-1.70$\\
NSC89&${}^1S_0$&$(0, -0.28)$&$(0, 1.47)$&$(2051.5, 0)$&$-2.48$\\
&${}^3S_1-{}^3D_1$&$(0, -0.45)$&$(0, 1.51)$&$(2046.8, 0)$&$-1.32$\\
ND&${}^1S_0$&$(0, -0.35)$&$(0, 1.49)$&$(2050.0, 0)$&$-1.83$\\
&${}^3S_1-{}^3D_1$&$(0, -0.35)$&$(0, 1.49)$&$(2050.0, 0)$&$-1.89$\\
NF&${}^1S_0$&$(0, -0.31)$&$(0, 1.48)$&$(2050.8, 0)$&$-2.19$\\
&${}^3S_1-{}^3D_1$&$(0, -0.36)$&$(0, 1.49)$&$(2049.7, 0)$&$-1.83$\\
 \end{tabular}
 \end{table}

\pagebreak

\noindent
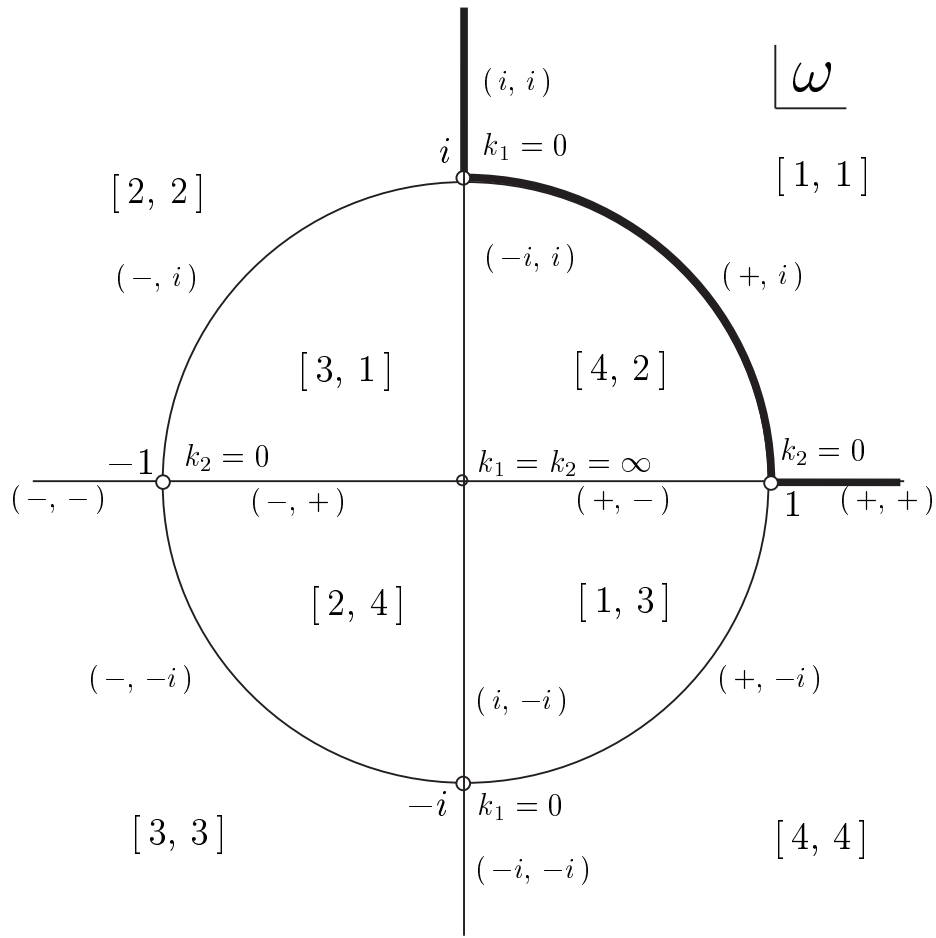
\begin{figure}
\caption{Complex $\omega$ plane into which the energy Riemann surface
is mapped.
The two numbers inside  the square brackets  indicate the quadrants
to which $q_1$ and $q_2$ belong, respectively. (The relation between the
energy $E$ and the momenta $q_1$ and $q_2$ is  given
by Eq.~(\ref{eq:4.1}).)
The parentheses show whether $q_1$ and $q_2$ are positive, negative, 
positive imaginary, or negative imaginary, respectively.     
The bold line expresses  the region where  bound or scattering states
exist if present.
\label{fig1}}
\end{figure}

\noindent
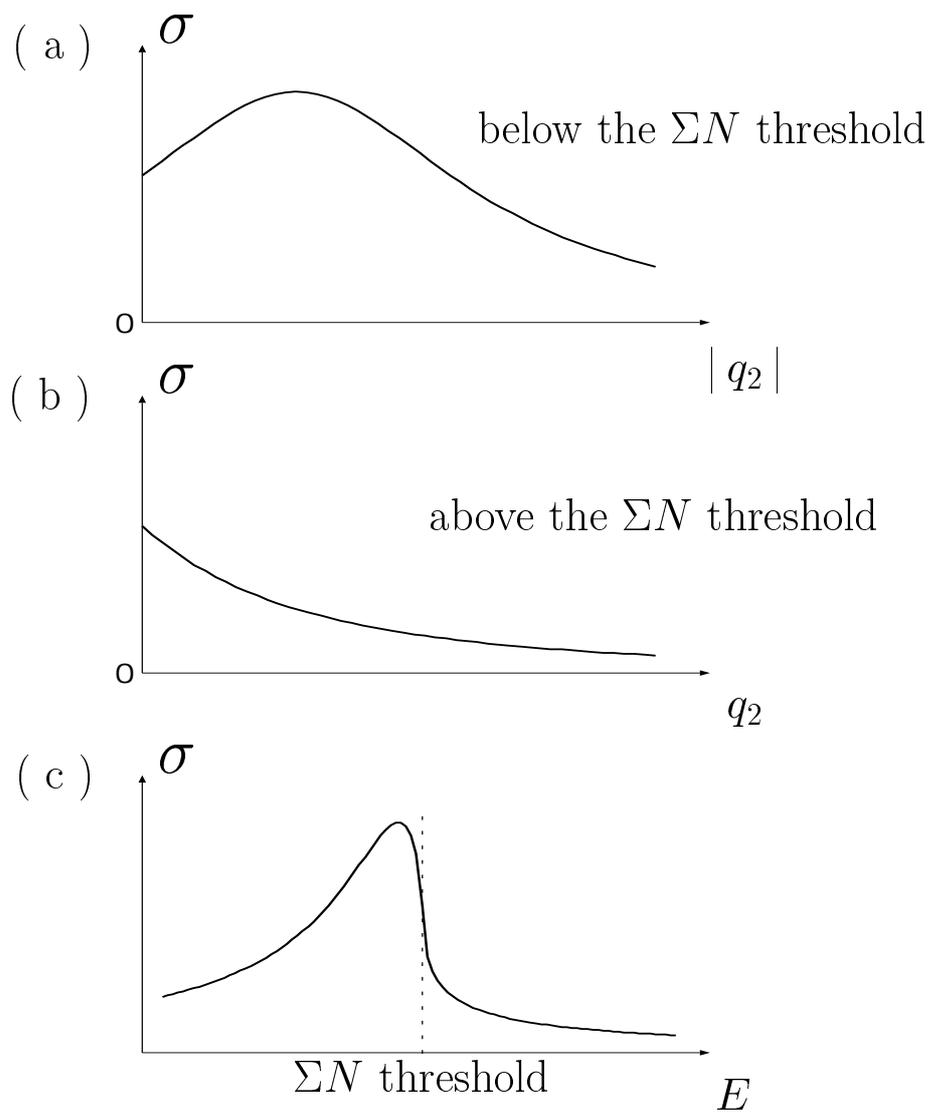
\begin{figure}
\caption{
Shape of the $\Lambda N$ elastic total  cross section
$\sigma$ around the $\Sigma N$ threshold in the case
a nearby $t$-matrix pole is located in the region [4,2]. (The pole
resides
in the 2nd quadrant of $q_2$.)
The cross sections $\sigma$ given by Eq.~(\ref{eq:4.10}) are plotted,
(a) as a function of $|q_2|$ below the threshold, (b) as a function 
of $q_2$ above the threshold, and (c) as a function of $E$.
 \label{fig2}}
\end{figure}

\noindent
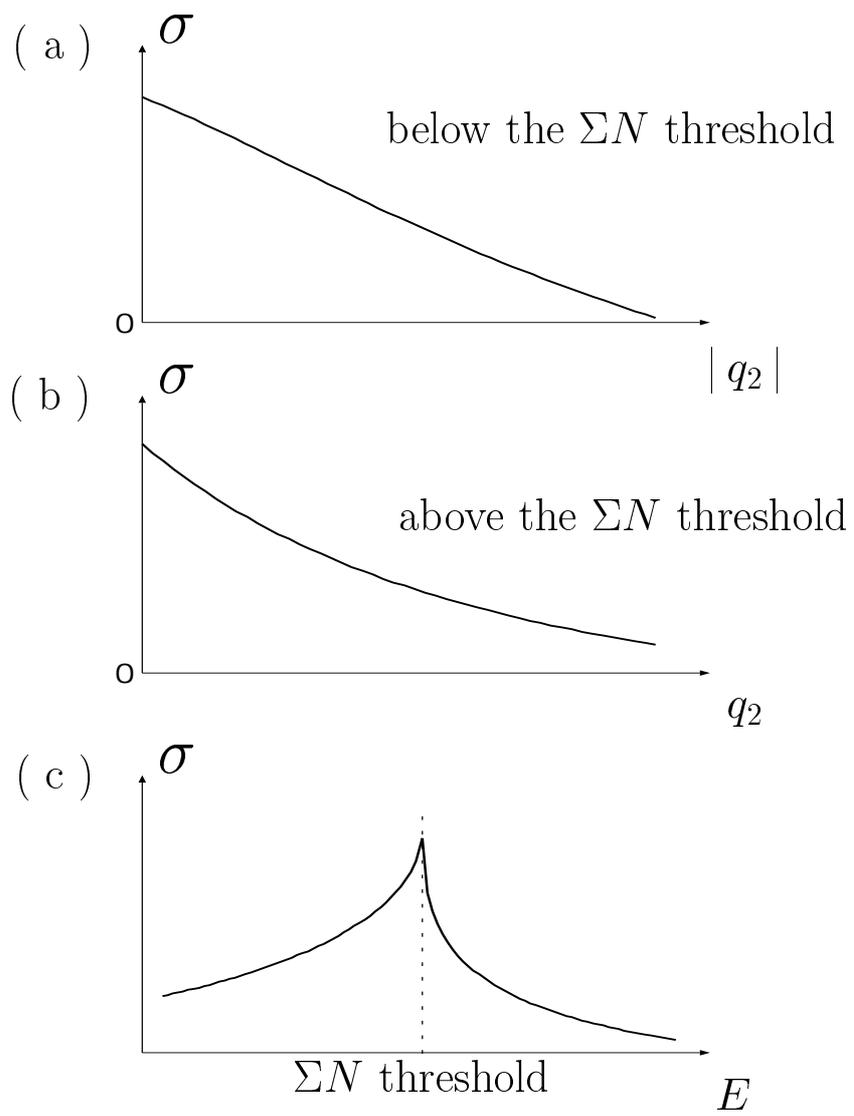
\begin{figure}
\caption{
Same as Fig.~2 but for the case
a $t$-matrix pole is located in the region [1,3]. (The pole resides in
the 3rd quadrant of $q_2$.)
 \label{fig3}}
\end{figure}

\noindent
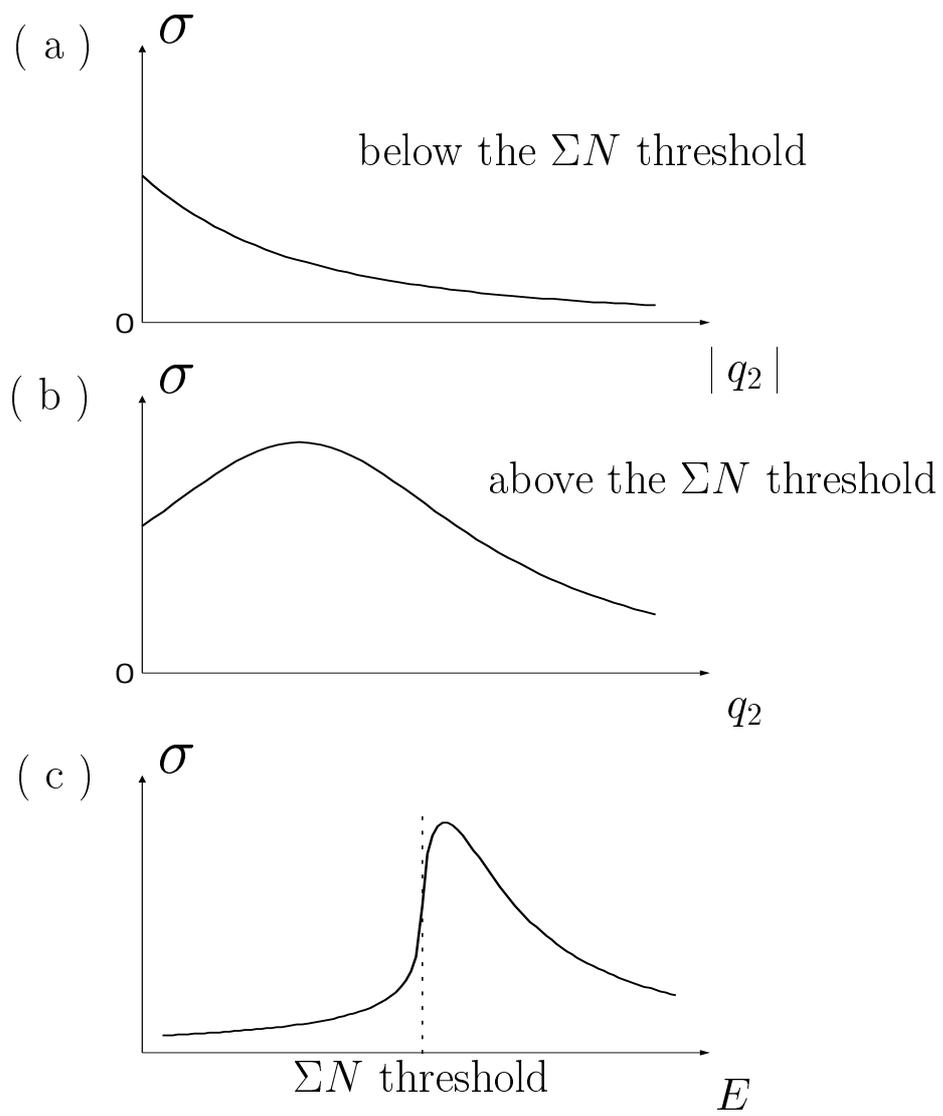
\begin{figure}
\caption{
Same as Fig.~2 but for the case
a $t$-matrix pole is located in the region [2,4]. (The pole resides in 
the 4th quadrant of $q_2$.)
\label{fig4}}
\end{figure}

\noindent
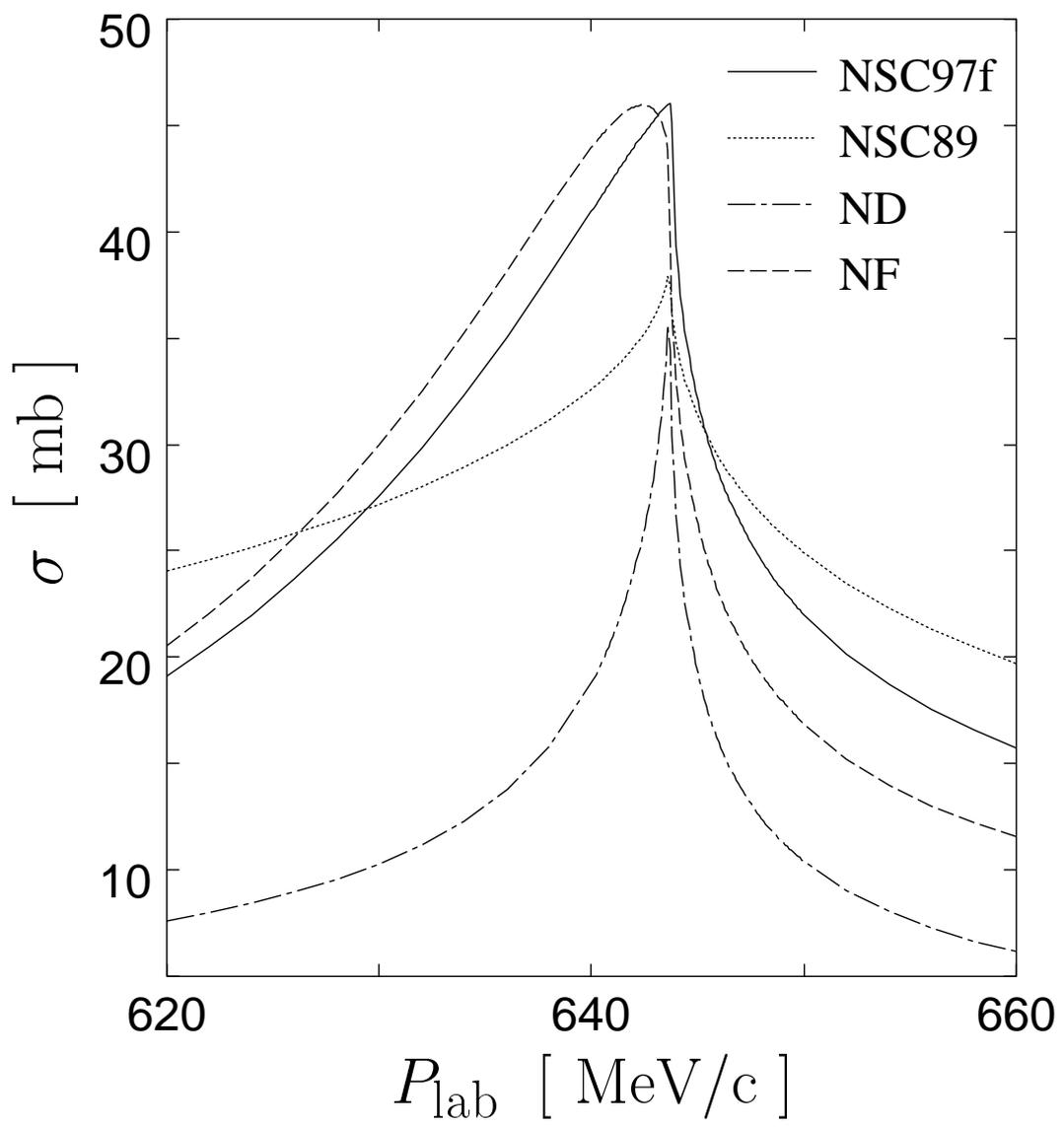
\begin{figure}
\caption{
$\Lambda N$ elastic total cross sections around the $\Sigma N$ 
threshold as a function of $\Lambda$ lab momentum.
Predictions by the force models of the Nijmegen group
NSC97f, NSC89, ND and NF are shown.
\label{fig5}}
\end{figure}

\noindent
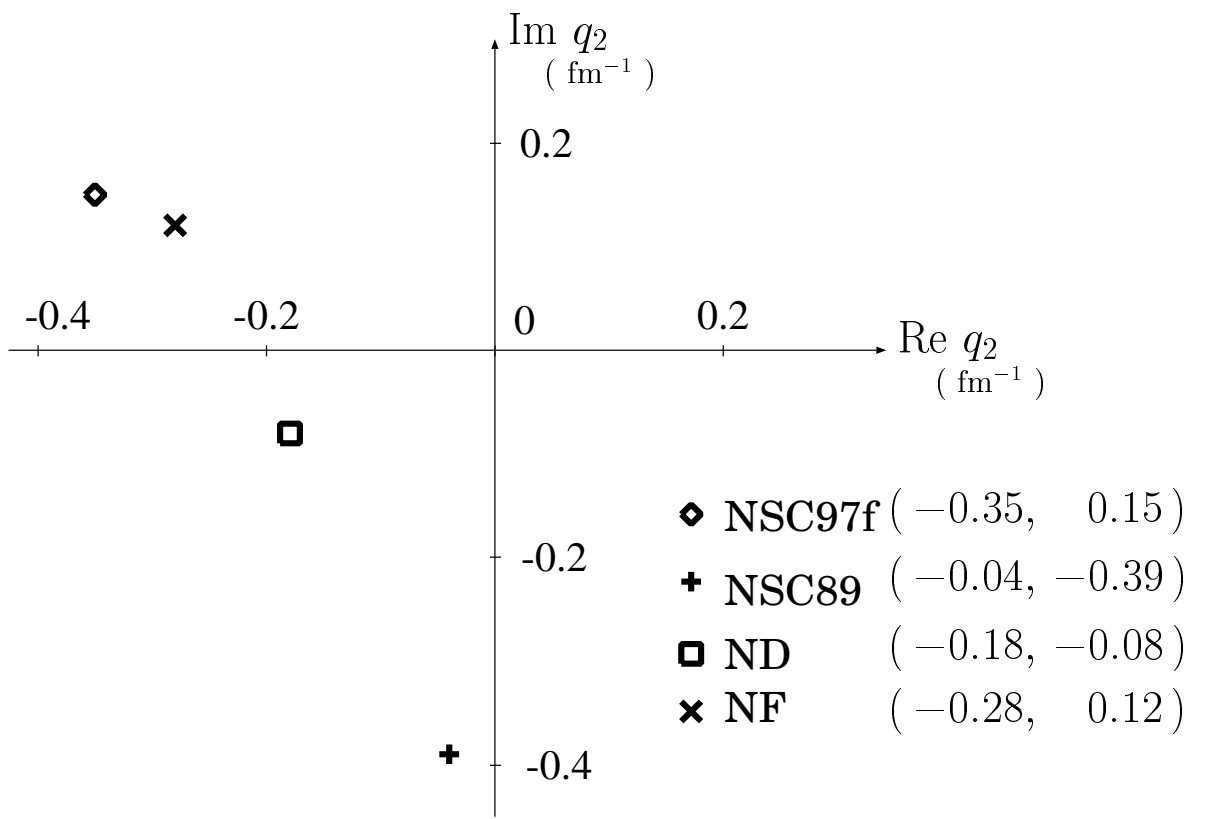
\begin{figure}
\caption{
Positions of the poles for the force models NSC97f, NSC89, ND and NF
in the complex $q_2$ plane.
\label{fig6}}
\end{figure}

\noindent
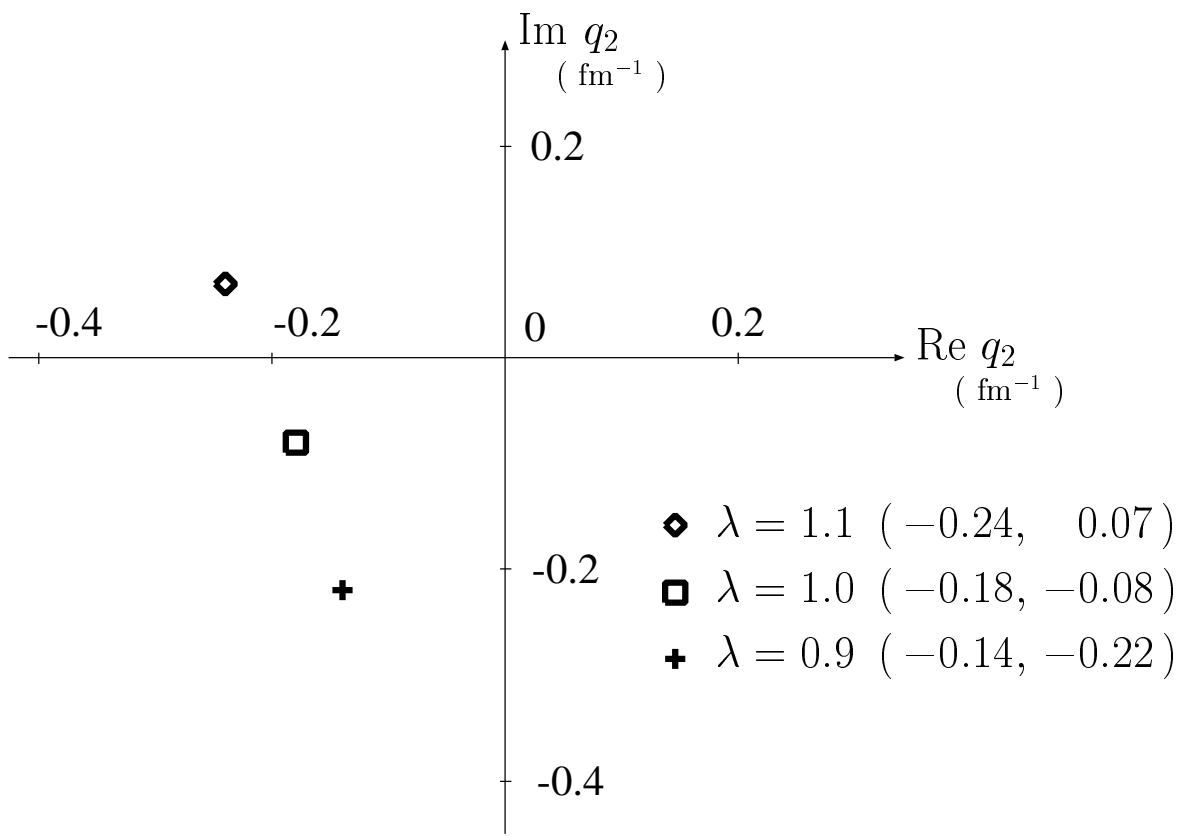
\begin{figure}
\caption{
Trajectory of the pole for the potential ND in the complex $q_2$ plane
with the multiplied  overall strength parameter  $\lambda$.
\label{fig7}}
\end{figure}

\end{document}